# A systematic comparison of structural, structural connectivity, and functional connectivity based thalamus parcellation techniques

Charles Iglehart[1], Martin Monti[2,3], Joshua Cain[2], Thomas Tourdias[4], Manojkumar Saranathan[1,5]

[1] Department of Electrical and Computer Engineering, University of Arizona, Tucson, Arizona, United States
[2] Department of Psychology, University of California Los Angeles, Los Angeles, California, United States
[3] Neurosurgery Brain Research Center, University of California Los Angeles, Los Angeles, California, United States
[4] Service de Neuroimagerie Diagnostique et Thérapeutique and INSERM U1215, Université de Bordeaux, Bordeaux, France
[5] Department of Medical Imaging, University of Arizona, Tucson, Arizona, United States

Corresponding author:
Charles Iglehart
ciglehart@email.arizona.edu
+1 773 746 5192
ORCID:  0000-0001-6665-104X

**Abstract**
The thalamus consists of several histologically and functionally distinct nuclei increasingly implicated in brain pathology and important for treatment, motivating the need for development of fast and accurate thalamic segmentation. The contrast between thalamic nuclei as well as between the thalamus and surrounding tissues is poor in T1 and T2 weighted magnetic resonance imaging (MRI), inhibiting efforts to date to segment the thalamus using standard clinical MRI. Automatic segmentation techniques have been developed to leverage thalamic features better captured by advanced MRI methods, including magnetization prepared rapid acquisition gradient echo (MP-RAGE) , diffusion tensor imaging (DTI), and resting state functional MRI (fMRI). Despite operating on fundamentally different image features, these methods claim a high degree of agreement with the Morel stereotactic atlas of the thalamus. However, no comparison has been undertaken to compare the results of these disparate segmentation methods. We have implemented state-of-the-art structural, diffusion, and functional imaging-based thalamus segmentation techniques and used them on a single set of subjects. We present the first systematic qualitative and quantitative comparison of these methods. We found that functional connectivity-based parcellation exhibited a close correspondence with structural parcellation on the basis of qualitative concordance with the Morel thalamic atlas as well as the quantitative measures of Dice scores and volumetric similarity index.

**Introduction**
The thalamus is a bilateral, subcortical deep brain structure that plays a critical role in numerous neurological processes, including consciousness, episodic memory, and attention. It consists of several histologically distinct nuclei also distinguishable by their global functional implications. The anterior nuclei figure prominently in episodic memory (Aggleton, 1999) and spatial processing (O'Mara, 2013); the mediodorsal (Md) nucleus has been shown to be involved in working memory (Watanabe et al. 2012); the lateral geniculate nucleus (LGN) exhibits a strong functional relationship with the visual cortex (Chen et al. 1998). Recent research has challenged the traditional pedagogic notion of the thalamus as a passive relay mechanism. For instance, the visual thalamic reticular and mediodorsal nuclei have been shown to dynamically modulate connectivity with the prefrontal cortex (Wimmer et al. 2015; Schmitt et al, 2017). The thalamic nuclei have also been shown to be selectively vulnerable to certain pathological conditions: changes in the limbic nuclei have been associated with the occurrence of Alzheimer's Disease (Braak and Braak 1991); volumetric changes in the Md and pulvinar (Pul) nuclei have been observed in persons afflicted by schizophrenia and schizotypal personality disorder (Andreasen 1997; Byne et al. 2001); degradation of the intralaminar nuclei has been associated with the progression of Parkinson's Disease (Henderson et al. 2000) as well as the depth of the impairment and recovery post severe brain injury (Schiff 2010); multiple nuclei bordering the third ventricle have shown selective vulnerability and atrophy in patients with multiple sclerosis (Planche et al. 2019). Recently, thalamic nuclei have been targeted for treatment of neurological disorders. For example, chronic high-frequency deep brain stimulation of the ventralis intermedius (VIM) nucleus has proven effective in the suppression of essential tremor (Benabid et al. 1991) and has shown the ability to enhance responsiveness in patients with severe disorders of consciousness (Schiff et al., 2007). Until recently, these nuclei could only be distinguished via histology or time consuming manual segmentation, but a quickly broadening array of noninvasive magnetic resonance imaging (MRI) based segmentation techniques are showing promise in rapidly, accurately, and automatically identifying thalamic nuclei *in vivo*.

MRI-based thalamus segmentation techniques developed to-date can be coarsely grouped into classes corresponding to different types of image acquisition, prominently among them structural, diffusion tensor imaging (DTI), and resting state functional MRI (fMRI). Structural image contrast arises from the variation of T1 or T2 relaxation parameters between different tissues; DTI operates upon signal variation corresponding to the restricted diffusion of molecular water within myelinated tissue; fMRI leverages the differential paramagnetism and diamagnetism of deoxyhemoglobin and oxyhemoglobin (respectively) to measure a blood oxygen level-dependent (BOLD) T2*-weighted signal. In general, variation of one these features in the thalamic nuclei need not necessarily correspond with the variation of any of the others. For instance, a region exhibiting a distinctive BOLD signal may be indistinguishable from surrounding tissues from the perspective of a structural acquisition. Thus, segmentation techniques based on differing acquisition types can be expected to produce different results.

Several parcellation methods based on structural MRI have been developed. Approaches using T1 and T2 weighted images were some of the first to be implemented (Deoni et al. 2005; Traynor et al. 2011), but these methods are inhibited by poor contrast between the intrathalamic nuclei as well as between the thalamus and surrounding tissues on these commonly collected image types. Sudyadhom et al (2009) developed a Fast Gray Matter Acquisition T1 Inversion Recovery (FGATIR) sequence for more reliably visualizing subcortical structures like the ventral intermediate nucleus for deep brain stimulation targets.

This sequence did allow for the localization of the entire thalamus but was not optimized to provide the intrathalamic contrast required for structural segmentation. Most structural imaging methods involve the use of Magnetization Prepared Rapid Acquisition Gradient Echo (MP-RAGE) or its variants. The method developed by Iglesias et al (2018) applied Bayesian methods to a probabilistic thalamus atlas derived from both ex-vivo histology and in-vivo MP-RAGE to segment the thalamus and was made publicly available as part of the Freesurfer processing platform. A recently developed technique called Thalamus Optimized Multi-Atlas Segmentation ('THOMAS', Su et al. 2019) uses White-matter-nulled (WMn) MP-RAGE (Saranathan et al. 2010) images to drive a multi-atlas segmentation technique that has shown excellent agreement with the Morel stereotactic anatomical atlas (Morel 1997).

The majority of methods published to date are diffusion-MRI based techniques. DTI can be used to examine microstructure both globally (e.g. through tractography and cortical connectivity analysis) and locally (via tensor analysis and scalar maps, including but not limited to fractional anisotropy (FA), mean apparent propagator (MAP), mean diffusivity (MD), etc.). Segmentation techniques based on global connectivity such as the tractography-based estimates of the cortical projections of particular thalamic nuclei (Behrens et al. 2003; Yamada et al. 2010; Jbabdi, Woolrich, and Behrens 2009) have shown to be reliable. However, they require manual delineation of the relevant cortical regions by an experienced neuroradiologist, which is time consuming. While these methods have demonstrated accuracy in identifying particular nuclear groups (i.e. the ventral nuclei), they have not appeared to reliably identify smaller nuclei. Locally, DTI can be leveraged in a variety of ways to provide voxel-level models of diffusion behavior. These models vary greatly in complexity and must be used judiciously to properly represent the structure of interest. Fitting second order tensors to the measured diffusion data assigns an ellipsoid to each voxel whose orientation and eccentricity correspond to the primary direction of diffusion and anisotropy, respectively (Basser, Mattiello, and LeBihan 1994). Clustering via dominant diffusion orientation (obtained via eigendecomposition of the diffusion tensors) has been used extensively to identify subregions of the thalamus (Unrath et al. 2008; Mang et al. 2012; V. Kumar, Mang, and Grodd 2015). Such methods are fast and simple to implement but can produce spatially disconnected regions owing to the fact that the clustering as implemented accounts exclusively for directional (and not spatial) proximity. Techniques incorporating measures of both spatial and tensorial similarity have been developed (Wiegell et al. 2003; Rittner et al. 2010), but the use of tensors as a diffusion model is ultimately and inherently limited by its inability to represent complex small scale fiber tract geometries such as intra-voxel bending, crossing, and twisting (Basser et al. 2000). Recent years have seen progressive refinement in the modeling used to represent small scale diffusion architecture. Q-Ball Imaging (QBI) (Tuch 2004) utilizes model independent spherical tomographic reconstruction to produce white matter fiber orientation distributions (FODs) capable of resolving fiber crossing. As more DTI directions are collected, at the expense of longer scan time, higher order QBI parametrizations can be performed, and finer scale structures can be modeled. The most consistent DTI-based segmentation developed to-date uses QBI-derived FOD parameters interior to the thalamus to identify specific nuclei (Battistella et al. 2017). In general, DTI's reliance on echo-planar imaging (EPI) results in lower resolution images relative to other acquisition types, typically on the order of 2-3 mm^3. In addition, the predominance of gray matter in the thalamus dictates a largely isotropic diffusion profile which has inhibited the performance of DTI-based segmentation techniques.

Analysis of resting state functional MRI (fMRI) has been used to explore functional relationships between cortical regions of interest (ROIs) and individual thalamic nuclei (Kumar et al 2017). Zhang et al (2008) proposed one of the first fMRI thalamic segmentation algorithms by correlating each thalamic voxel against a set of cortical ROIs delineated by their distinct functional roles. This study showed that each thalamus voxel correlated strongly with only one region, suggesting strong resting-state functional specialization. Ji et al (2016) built upon this by using several temporally independent thalamocortical states with generally closer correlation with the Morel atlas, but this method required a large number of subjects and intensive processing. The method proposed in (van Oort et al. 2018) extends the use of fMRI for segmentation of cortical and subcortical structures including the thalamus. This technique leverages instantaneous temporal correlations in the BOLD signal to identify functionally distinct subregions within a ROI. Instantaneous connectivity parcellation (ICP) techniques are used to partition the thalamus via subspace projection. This method also employs a data-driven technique for identifying an optimal number of segments for the final segmentation.

Methods using each image acquisition class claim high measures of agreement with the same anatomical atlases (typically the Morel atlas), yet the underlying acquisition types measure fundamentally different features. This suggests that an examination is warranted to determine how parcellation methods predicated on differing acquisition classes compare and contrast both qualitatively and quantitatively. Furthermore, such a comparison might facilitate multimodal segmentation, if the techniques provided similar and/or complementary information. Towards that end, this study presents the first systematic comparison of structural, structural-connectivity, and functional-connectivity based thalamus segmentation techniques run on the same set of subjects.

**Methods**

*MRI acquisition*

Multiple image types were collected for 18 healthy individuals who provided prior informed consent following the UCLA IRB approved procedures. All images were obtained using the Siemens 3 Tesla MAGNETOM Prisma fit MRI machine (Erlangen, Germany) housed at the Staglin Center for Cognitive Neuroscience at UCLA. Pulse sequence parameters for the different sequences are as follows- T1-weighted MP-RAGE: 192 sagittal slices, TR/TE 2,000/2.52 ms, 12° flip angle, 1mm isovoxel resolution, FoV 256 mm, generalized autocalibrating partially parallel acquisitions ("GRAPPA", Griswold (2002)) acceleration factor 2; White-matter nulled (WMn) MP-RAGE: 160 axial slices, TR/TE 4,000/3.75 ms, inversion time 500 ms, 7° flip angle, 1mm isoxovel resolution, FoV 256 mm, with GRAPPA acceleration factor 2; Resting state BOLD EPI: 54 axial slices, TR/TE 700/33 ms, 70° flip angle, 2.5 mm isovoxel, using a simultaneous multislice ("SMS", Barth (2016)) acquisition with acceleration factor 6, for a total of 860 volumes; DTI: 50 axial slices, 64 directions, with three b-values of 0 and one 1000 s/mm^2, TR/TE 7,000/93 ms, with 2.0 mm isovoxel resolution, FoV 190 mm.

After a careful review of the literature, three recently developed and promising algorithms for structural, DTI, and fMRI based parcellation were chosen for implementation. THOMAS (Su et al. 2019) was selected for structural segmentation; a FOD-driven method (Battistella et al. 2017) was used for DTI-based segmentation; rs-fMRI segmentation was performed via ICP (van Oort et al. 2018). These methods were chosen for their consistency, reproducibility, and their ability to produce a sufficient number of spatially contiguous labels to facilitate comparison with the Morel atlas. They are briefly described below.

*Structural segmentation method*

Structural segmentation was performed with THOMAS (Su et al. 2019), which uses multiple image registration steps and multi-atlas label fusion to effect thalamic segmentation as depicted in Figure 1. It uses a set of 20 prior WMn volumes (multi atlas) manually segmented by a trained neuroradiologist into 11 distinct nuclei using the Morel stereotactic atlas as a reference. The atlas prior images were also nonlinearly warped to a common space and averaged to produce a high SNR template image, with the corresponding prior-template transformations stored to reduce computation time in subsequent steps. To perform thalamic segmentation on an input image, the input image was first cropped automatically to encompass both thalami and then nonlinearly warped to a cropped template volume. By concatenating this warp with the pre-computed prior-template warps, thalamic nuclei labels from each manually segmented prior were warped to the input space. The final labels were determined via a joint label fusion process called PICSL-MALF (Wang and Yushkevich 2013), wherein each prior label's contribution to the input segmentation is weighted by that prior WMn's similarity to the input WMn at each voxel.

*DTI segmentation method*

DTI segmentation was performed via a modified version of the technique described in (Battistella et al. 2017). This method attempts to leverage local diffusion characteristics within the thalamus to inform the segmentation process. Figure 2 gives an overview of the algorithm architecture. DWI denoising, eddy current, and EPI distortion corrections were run using Mrtrix 3.0 'dwidenoise' and 'dwipreproc' utilities. All image registration and transformation was performed via Advanced Normalization Tools (ANTs, ("ANTs by Stnava" n.d.)). Image segmentation code was written in both Python (Python Software Foundation, https://www.python.org/) and Matlab (The MathWorks, Inc., Natick, Massachusetts, United States). Battistella et al perform thalamus masking through cortical and subcortical parcellation in Freesurfer and SPM-8. While reliable, these parcellation are slow and computationally intensive. Instead, we leveraged the WMn MP-RAGE averaged template developed for THOMAS and a template-based registration method to mask off the input image thalamus with considerable time and computational savings (Figure 2). Furthermore it aided the comparison between structural and DTI-based segmentation that was conducted from the same global thalamus volume. Each individual subject's WMn MP-RAGE volume was registered to the THOMAS WMn MP-RAGE template space via the above-described method. The template thalamus was then nonlinearly warped to the input WMn space to obtain the input thalamus mask. This mask was transferred to the input DTI space via another series of transformations. The remainder of the segmentation was done as described in Battistella et al. FODs were computed at each DTI voxel via the FSL '*qboot*' utility, which implements the Q-Ball Imaging algorithm (Tuch, 2004). QBI was used with an order 6 spherical harmonic basis to determine 28 FOD coefficients per voxel. The FODs were interpolated to 1mm isotropic resolution. K-means clustering with a modified distance metric incorporating both Euclidean voxel and spherical harmonic (SH) coefficient proximity produced the final segmentation. To ensure that voxel and SH distances contributed equally to the metric, SH coefficients were scaled by an empirically determined factor of 100. To account for k-means' sensitivity to initial seeding, the centroids of the labels resulting from 5000 k-means runs were used to seed the final segmentation. K-means also requires the number of output clusters be specified in advance of the analysis. Seven clusters were used in this study in accordance with the source literature (Battistella et al. 2017).

*Resting state fMRI segmentation method*

Resting state fMRI segmentation was performed using Instantaneous Connectivity Parcellation (ICP; van Oort et al. 2018; V. J. Kumar et al. 2017). Prior to applying this method, resting state blood oxygenation level dependent (BOLD) data were first preprocessed using FSL (Smith et al. 2004) to remove 4 initial volumes, apply slice time correction to account for the sequential (bottom-up) acquisition of 2D slices at each volume, perform rigid-body realignment to account for between-volume motion, and spatially smooth with a Gaussian kernel of 3.5 mm FWHM. It was then temporally smooth with a high-pass filter of 0.01 Hz only, since prior work has shown that resting-state information can be found at much higher frequencies than can be observed when a conventional 0.1 Hz low-pass filter is also used [i.e., to create a band-pass filter; (Wu et al. 2008; Lee et al. 2013). Finally, motion-related artifacts were minimized using the aCompCor50 procedure with 24 motion parameters (Muschelli et al. 2014) and with additional regressors coding for individual high-motion volumes (i.e., 'spike regression;' (Satterthwaite et al. 2013)) as defined by the root mean square difference between a volume and the mid-point volume reference exceeding the outlier threshold of the 75th percentile plus 1.5 times the inter-quartile range (as implemented in fsl_motion_outliers). The residuals forming this nuisance regression were then submitted to the ICP pipeline, which was implemented with an in-house tcsh script combining tools from FSL, Matlab, and AFNI. Following the procedure described by van Oort et al, the ICP was implemented in the following steps applied to each subject and hemisphere separately (see figure 3): (i) a thalamic "initialization mask" was created by segmenting the individual T1-weighted MP-RAGE data using FSL FIRST (Patenaude et al. 2011), as implemented in the fsl_anat (https://fsl.fmrib.ox.ac.uk/fsl/fslwiki/fsl_anat) pipeline, and projected into BOLD space; (ii) the average time-course of all voxels within the initialization mask (for each hemisphere separately) was extracted and standardized; (iii) the time-series of each voxel within the initialization mask were also standardized (individually); (iv) the standardized vector obtained in (ii) and the standardized voxel time-series obtained in (iii) were then multiplied element-wise, for each voxel separately, thereby creating the "unfolded" time-series of instantaneous correlations [cf., van Oort et al., 2018] which were then transformed into MNI template space with a 12 degrees of freedom affine transformation (using FSL FLIRT (Jenkinson and Smith 2001) ). These "unfolded" time-series differ from the original voxel time-series in that they amplify transient events between related time courses, thereby helping to identify instantaneous correlations between voxels [van Oort et al., 2018]. Finally, (v) the unfolded time-series were entered into a group ICA analysis using a temporal concatenation model [as implemented in MELODIC; (Beckmann and Smith 2004) ] in order to detect components with coherent spatial topography across individuals (but without requiring them to have similar time-courses across individuals). In order to align the parcellation output across methods, the group ICA enforced a desired solution dimensionality of 30 components, which has been shown in a large cross-validation study to yield the most reproducible results [van Oort et al., 2018]. In addition, a dual regression (Beckman et al., 2009; Nickerson et al., 2017) approach was employed in order to obtain single-subject thalamic ICPs. Specifically, for each subject, the group-average set of parcellation were regressed (as spatial regressors in a multiple regression) into each subject's time-series of instantaneous correlations, thereby generating one subject-specific time-series per group-level parcel. Next, these time-series were regressed (as temporal regressors) into the same 4D dataset, resulting in a set of subject-specific spatial maps, one per group-level spatial map.

In accordance with the method proposed by van Oort et al, the fMRI segmentation was designed to output 30 clusters. This design parameter acting in combination with voxel size (2.5 mm isotropic) and the relatively small volume of the thalamus dictated that many of the of the resulting clusters consisted of only a few voxels and were thus of limited anatomical relevance. To facilitate comparative analysis, the 30 fMRI clusters were regrouped. Group level maximum probability maps (see description below) were computed in MNI space for both THOMAS and fMRI. Overlap was computed between each of the constituent labels of the corresponding maps. Each of the 30 fMRI maximum probability labels was mapped to one of the 11 THOMAS maximum probability labels on the basis of overlap. This mapping was subsequently applied on an individual level to obtain regrouped individual fMRI segmentation with 11 labels.

*Post processing and analysis*

The output segmentation were registered and warped (using nearest neighbor interpolation) to the MNI305 T1 weighted statistical atlas for the purpose of analysis. Probability maps were generated for each parcellation technique. Created in the common MNI305 space, these maps can be used to gain insight into average, group-level segmentation characteristics (Najdenovska et al. 2018). Two types of probability maps were created in the MNI305 space - weighted and maximum. Weighted probability maps were generated by assigning a color to each thalamus voxel representative of the relative proportions of labels present in that voxel. These maps can be used as a means of evaluating the spatial consistency of a parcellation in common space. Diffuse transitions between regions on the map can be indicative of spatial inconsistencies. Maximum probability mapping assigned colors dictated by the label most commonly present within each voxel. These maps were used to regroup the fMRI parcellation as well as to explore group-level segmentation characteristics.

In addition to using probability maps to gauge consistency, label centroid maps were also created. For each technique, the spatial centroid of each individual label in MNI305 space was computed. Centroids for all individuals were overlaid on images of the thalamic region in all three planes. Each centroid label was assigned a unique color code. Spatial clustering of the label centroids provided a means of evaluating each method for consistency – a tighter clustering of centroids for a particular label in MNI space is indicative of higher spatial consistency.

Two quantitative metrics were calculated to compare the three methods across all subjects: Dice and Volumetric Similarity Index (VSI). For two labels $L_1$ and $L_2$, Dice and VSI are defined:

$$Dice(L_1, L_2) = 2 \frac{\|L_1 \cap L_2\|}{\|L_1\| + \|L_2\|}$$

$$VSI(L_1, L_2) = 1 - \frac{|\ \|L_1\| - \|L_2\|\ |}{\|L_1\| + \|L_2\|}$$

where $\|\cdot\|$ measures the number of voxels in a label, the intersection operation $L_1 \cap L_2$ gives the set of voxels common to both $L_1$ and $L_2$, and $|\cdot|$ is the absolute value operation. Individual Dice and VSI were averaged to obtain group-level statistical characterizations.

**Results**

The fMRI label regrouping process is outlined in figure 4. The initial fMRI segmentation (using 30 clusters) produced numerous labels comprising only a few voxels and were thus of dubious anatomical significance. Upon regrouping to 11 labels, the spurious clusters were reassigned to larger groups that simplified qualitative evaluation against the Morel atlas.

The three parcellation methods were used to segment the right and left thalamus for 18 subjects. Figure 5 shows left and right thalamus segmentation results using all 3 methods for two subjects for the same axial slice. From visual inspection, the structural based method THOMAS exhibits the highest degree of agreement with the Morel atlas. It clearly delineates several major nuclei: the anterior ventral, the ventral nuclei (VPL, VLa, VLP), the Pulvinar, and the mediodorsal.

Since the DTI-based method used only seven clusters to segment the thalamus, it does not offer fine scale resolution. As shown in figure 5, it clusters the AV, VA, and parts of the VLa nuclei into a single anterior label. It also tended to split the Pulvinar region roughly equally into lateral, medial, and superior sections. The nucleus it appears to identify most reliably is the mediodorsal nucleus.

The fMRI segmentation appears to generate highly consistent divisions of the thalamus after regrouping. It compares favorably with the Morel atlas and performed especially well in discriminating larger nuclei such as the mediodorsal, ventral, and pulvinar. However, it also appears to overestimate the sizes of smaller structures such as the anterior ventral (AV) and center median (CM) nuclei. Also, some smaller nuclei as the lateral and medial geniculate (LGN and MGN) and habenular (Hb) were not identified due to exclusion of the corresponding voxels during the masking process.

Weighted probability maps for each method in axial, coronal, and sagittal planes are shown in figure 6. The weighted DTI maps are the most blurred, suggesting a higher degree of spatial variation both within and between label groups. The weighted structural and fMRI maps are less blurred, suggesting greater spatial consistency. The coronal and sagittal views show that the structural maps clearly delineate the LGN and MGN. The DTI map also includes these regions in its thalamic mask, but incorporates these small peripheral nuclei into larger labels comprising the pulvinar region.

Figure 7 shows centroids for all three methods visualized in all three planes. Again, structural and fMRI label centroids are tightly clustered, suggesting consistency, while DTI labels are evidently more variable. It is worth noting that spatial consistency is likely to decrease with the number of labels used.

Tables 1 and 2 summarize group-level Dice and VSI statistics between DTI, fMRI, and structural segmentation for eleven nuclei of interest (similarity statistics between DTI and fMRI are given in supplemental material). Scores for right and left thalamus are presented independently. Between structural and fMRI parcellation, mean Dice scores are highest in the pulvinar (0.61/0.65 left/right) and mediodorsal

(0.57/0.51 left/right) regions. DTI and structural also show highest mean Dice similarity in the large nuclei: 0.60/0.57 in the mediodorsal and 0.50/0.48 in the pulvinar. Among some smaller nuclei, fMRI demonstrated higher Dice with structural: 0.24/0.24 in the AV, 0.44/0.28 in the CM (as opposed to 0.14/0.15 and 0.17/0.14 for DTI). However, fMRI masking excluded some voxels peripheral to the thalamus corresponding to small nuclei as the LGN and Hb, resulting in no Dice value between fMRI and structural for those nuclei. VSI was notably higher for fMRI: 7 of the 11 nuclei reported a VSI at 0.9 or higher (corresponding to the 7 largest nuclei by mean volume as measured by structural); DTI exhibited VSI in excess of 0.9 for two nuclei.

| Nucleus | Dice (L) | Dice (L) | Dice (R) | Dice (R) |
|---|---|---|---|---|
| VL | 0.48 ± 0.06 | 0.51 ± 0.06 | 0.46 ± 0.08 | 0.43 ± 0.06 |
| Pul | 0.50 ± 0.06 | 0.61 ± 0.11 | 0.48 ± 0.06 | 0.65 ± 0.09 |
| Md | 0.60 ± 0.06 | 0.57 ± 0.11 | 0.57 ± 0.09 | 0.51 ± 0.10 |
| VPL | 0.33 ± 0.08 | 0.35 ± 0.07 | 0.38 ± 0.06 | 0.43 ± 0.10 |
| VA | 0.44 ± 0.05 | 0.37 ± 0.07 | 0.42 ± 0.05 | 0.50 ± 0.11 |
| AV | 0.14 ± 0.03 | 0.24 ± 0.09 | 0.15 ± 0.04 | 0.24 ± 0.08 |
| CM | 0.17 ± 0.02 | 0.44 ± 0.11 | 0.14 ± 0.03 | 0.28 ± 0.04 |
| LGN | 0.28 ± 0.04 | - | 0.31 ± 0.06 | - |
| VLa | 0.12 ± 0.04 | 0.20 ± 0.08 | 0.10 ± 0.04 | 0.28 ± 0.08 |
| MGN | 0.13 ± 0.04 | 0.02 ± 0.01 | 0.12 ± 0.04 | 0.16 ± 0.05 |
| Hb | 0.03 ± 0.01 | - | 0.03 ± 0.01 | - |

**Table 1** Mean +/- standard deviation Dice scores between left and right DTI and fMRI parcellation compared with structural parcellation across 18 subjects for 11 nuclei. Nuclei are arranged in descending order of average label volume.

| Nucleus | VSI (L) | VSI (L) | VSI (R) | VSI (R) |
|---|---|---|---|---|
| VL | 0.97 ± 0.02 | 0.95 ± 0.07 | 0.97 ± 0.02 | 0.92 ± 0.14 |
| Pul | 0.81 ± 0.07 | 0.94 ± 0.11 | 0.88 ± 0.05 | 0.93 ± 0.12 |
| Md | 0.97 ± 0.03 | 0.90 ± 0.06 | 0.98 ± 0.03 | 0.93 ± 0.11 |
| VPL | 0.71 ± 0.06 | 0.95 ± 0.04 | 0.75 ± 0.08 | 0.90 ± 0.06 |
| VA | 0.69 ± 0.06 | 0.94 ± 0.04 | 0.80 ± 0.08 | 0.93 ± 0.06 |
| AV | 0.41 ± 0.07 | 0.96 ± 0.04 | 0.48 ± 0.08 | 0.96 ± 0.03 |
| CM | 0.35 ± 0.12 | 0.90 ± 0.07 | 0.36 ± 0.04 | 0.91 ± 0.07 |
| LGN | 0.39 ± 0.10 | - | 0.40 ± 0.06 | - |
| VLa | 0.28 ± 0.18 | 0.73 ± 0.12 | 0.26 ± 0.05 | 0.73 ± 0.09 |
| MGN | 0.26 ± 0.19 | 0.70 ± 0.08 | 0.25 ± 0.03 | 0.72 ± 0.08 |
| Hb | 0.11 ± 0.12 | - | 0.09 ± 0.02 | - |

**Table 2** Mean +/- standard deviation VSI scores between left and right DTI and fMRI parcellation compared with structural parcellation across 18 subjects for 11 nuclei. Nuclei are arranged in descending order of average label volume.

| Nucleus | Abbreviation |
|---|---|
| Anterior ventral | AV |

| Center median | CM |
|:---:|:---:|
| Habenular | Hb |
| Lateral geniculate | LGN |
| Medial geniculate | MGN |
| Mediodorsal | Md |
| Pulvinar | Pul |
| Ventral anterior | VA |
| Ventral lateral | VL |
| Ventral lateral anterior | VLa |
| Ventral posterior lateral | VPl |

**Table 3** Index of abbreviations for 11 thalamic nuclei.

**Discussion**

Reliable and automatic thalamic nuclei segmentation is of considerable value across a broad array of clinical and research applications. While many segmentation techniques have been proposed, there exists insufficient analysis of the relative merits of these methods which are dependent on acquisition types. Towards this end, we implemented and compared what we consider the state-of-the-art structural, diffusion tensor, and resting state functional MRI-based thalamic nuclei segmentation methods on a single set of 18 subjects. Both individual segmentation and group level probability maps for each method were largely consistent with those detailed in the corresponding source literature.

In the absence of ground truth manual segmentation, THOMAS was used as a proxy. It had the highest spatial resolution of the three acquisition techniques and has been validated against manual segmentation guided by the Morel atlas (Su et al 2019). DTI and fMRI parcellation show different correspondence patterns with THOMAS. On the basis of Dice scores, DTI segmentation identifies the MD nucleus and some smaller nuclei (LGN, Hb) more reliably than fMRI. Otherwise, fMRI shows a largely higher degree of agreement with nuclei as delineated by structural segmentation. The VSI between fMRI and THOMAS is high for several large nuclei, further underscoring concordance.

The Morel atlas was constructed in part through evaluation of immunostaining for three calcium binding proteins commonly found in the thalamus: parvalbumin (PV), calbindin (CB), and calretinin (CR). Predominance of these proteins was observed to correlate with distinct regions of functional significance. For instance, differing somatosensory pathways were characterized by contrasting PV and CB concentrations. The twenty priors manually labeled used in THOMAS were delineated manually with the Morel atlas as a guide. Thus, the consistent agreement between THOMAS and the regrouped fMRI outputs could indicate that the fMRI technique is capable of producing labels coincident with structural subregions of the thalamus.

The computed probability maps support observations made from individual segmentation: 1) DTI tends to identify the Md nucleus better than it does other nuclei, it conglomerates the AV and VA nuclei, and splits the pulvinar into three parts (medial, lateral, superior); 2) THOMAS is most reflective of anatomy per the Morel atlas: it partitions the anterior and ventral nuclei and properly outlines the pulvinar and mediodorsal regions. However, it does not resolve the pulvinar into constituent subdivisions; 3) fMRI produces many small but highly consistent parcels, some of which could correspond to smaller subunits such as the divisions of the pulvinar.

As noted earlier, fMRI largely fails to distinguish smaller nuclei such as the Hb, while DTI shows a greater but still limited capacity to do so. This could be due in part to the size of these nuclei in relation to image resolution (1.97x1.97x3.0 mm for DTI, 2.5mm isotropic for fMRI), which are typically only a few voxels in extent. Higher resolution DTI and fMRI could enhance the fidelity of these nuclei. The use of techniques such as MUSE (Chen et al. 2013) can enable higher resolution (sub mm) DTI via multi-shot EPI. Alternatively, superresolution techniques have been developed to synthesize high-resolution DTI using multiple low-resolution images (Peled and Yeshurun 2001) but are susceptible to subject motion.

Local DTI segmentation is advantageous relative to global methods in that long-distance fiber tracking may be unreliable in patients with pathology such tumors or white matter or gray matter lesions, which will disrupt the tractography process. However, exclusive reliance upon local diffusion behavior in analyzing thalamic substructure may also be of limited utility. The thalamus is a predominantly grey matter structure with low fractional anisotropy, and so the degree to which the complex FODs used by this method are representative of true architecture is questionable. Additional examination would be warranted to explore the effect of varying DTI acquisition parameters upon the final segmentation product. A technique leveraging properties of multiple shell collections (Multi-shell, multi-tissue constrained spherical deconvolution, or MSMT CSD) has been developed that allows for the estimation and isolation of WM-only FODs; segmentation based upon these may be more reflective of relevant myelinated structure. Of course, the predominance of grey matter in the thalamus could call into question the utility of WM-only FODs. Collecting a larger number of diffusion directions during DTI acquisition (and using higher b-values) would permit the use of higher order spherical harmonic basis functions in FOD modeling schema, which in turn could allow for improved identification of smaller scale structures in the segmentation process. Susceptibility weighted images acquired at 7T have been shown promise as a means of providing supplementary information about the thalamus for DTI collected on the same individual (Abosch et al. 2010). Deep learning could be used to exploit unseen relationships between different acquisition modes to produce enhanced segmentation. A multi-modal, deep-learning based technique has been developed for the purposes of cortical parcellation (Glasser et al. 2016). This sort of architecture could be trained to leverage relationships between structural, structural connectivity, and functional connectivity localized to the thalamic nuclei. The DTI-based method we chose accounts for the inherent sensitivity of k-means clustering to starting centroid choices using a clever data-driven initialization. However, an evaluation of probability maps and nucleus centroids shows that it exhibits a higher degree of intra-subject variability than THOMAS and fMRI. Ideally, this comparison of methods would be performed upon a larger number of subjects (than the 18 used in this study) to strengthen group-level statistical conclusions.

It should be noted that this analysis had several limitations. The fMRI method did not assign labels to the LGN and MGN regions in contrast with structural and DTI based parcellation. This is due to the truncated thalamic mask applied to the fMRI data rather than a defect in the parcellation technique itself. This analysis would have been better performed using the same thalamic mask across all techniques. The regrouping of fMRI parcellation on the basis of group-level probabilistic relationships with structural could reasonably be called into question. The use of structural as ground truth is less than ideal. Ideally, manually segmented truth should have been used for the purposes of evaluation and statistical comparison but was beyond the scope of this study. The manner in which the number of output thalamic subdivisions was chosen differs between methods. THOMAS produces 11 nuclei in accordance with the Morel atlas; the fMRI method used a data-driven split-half reproducibility to settle upon 30 clusters. (The fMRI data was also used to produce an 11 nucleus segmentation via ICP, but these clusters were less stable than those produced by the regrouping method. Examples of individual segmentation and label probability maps for the 11 nucleus ICP fMRI segmentation are provided in supplemental material). DTI segmentation divides the thalamus into 7 regions incidentally in accordance with Behrens' global segmentation method. However, the clustering used in that study corresponded to the number of pre-segmented cortical targets used (Behrens et al, 2003). While the data-driven methodology used in the fMRI technique to choose the number of segmentation nuclei is preferable to a more arbitrary user-supplied parameter, it is unclear whether the 30 subdivisions it generates are best representative of the actual functional architecture of the thalamus. Despite its design, THOMAS did not resolve all finer scale nuclei from the Morel atlas; for instance, it does not identify the anterior or medial pulvinar, rather identifying a single pulvinar region. Thus, further study would be justified to investigate the optimal number of segmentation clusters for each method.

**Conclusion**
We have implemented a systematic comparison of structural, DTI, and fMRI-based thalamus parcellation techniques on a single set of subjects. Each method exhibited strengths and weaknesses that could inform their future use. Structural parcellation bears the strongest resemblance to the Morel stereotactic atlas. Structural and fMRI based methods were more spatially consistent than DTI. Structural and fMRI segmentation bore a closer statistical similarity to one another relative to DTI. Each method performed best in identifying larger nuclei.

**Figures**

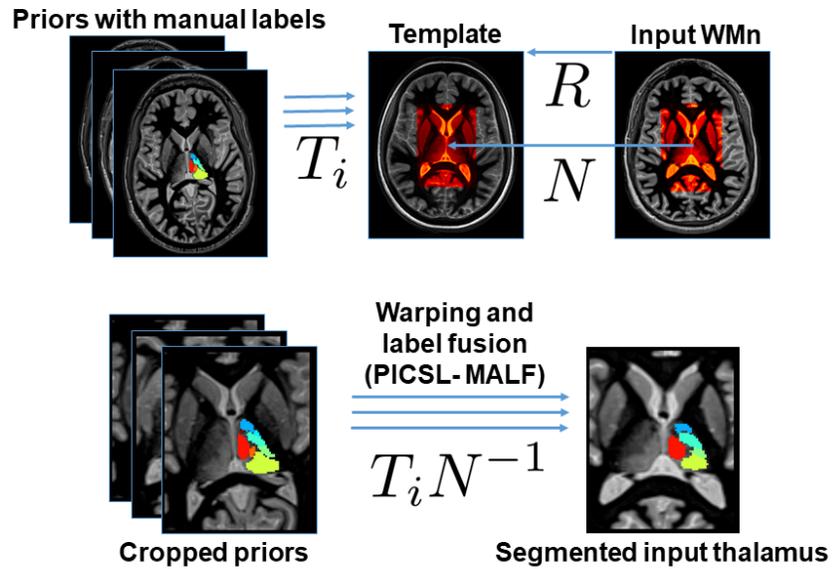

**Fig. 1** Structural MRI-based segmentation method (THOMAS). (i) Manually segmented prior volumes are nonlinearly registered to a template volume with the resulting transformations stored for future use; (ii) the entire input volume is rigidly registered to the template; (iii) a cropped region of the input surrounding the thalamus is nonlinearly warped to the template; (iv) the prior thalamic labels are warped first to the template space and then to the individual input space; (v) the segmented input thalamus is obtained via label fusion of the warped priors.

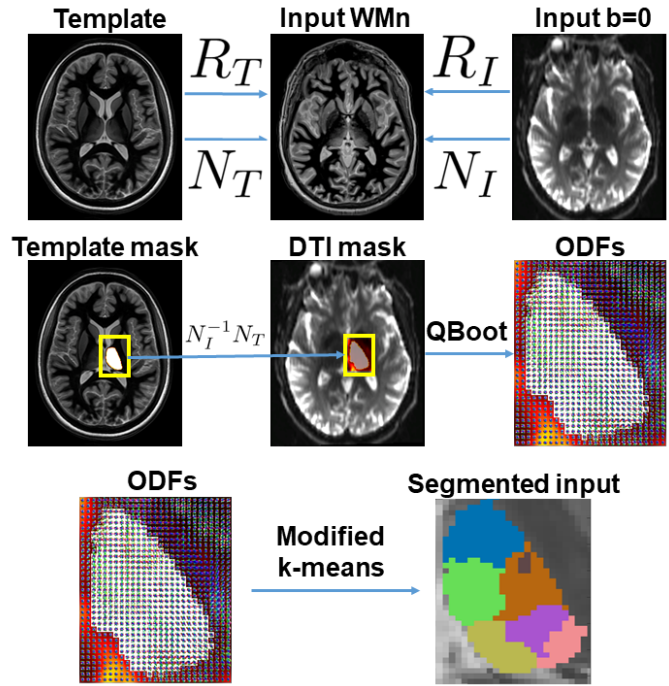

**Fig. 2** DTI-based segmentation method. (i) An individual thalamus mask is estimated via the same multi-stage registration process used in THOMAS; (ii) the mask is used to isolate thalamus voxels within the DTI; (iii) order 6 (28 coefficient) FODs are generated at each voxel in the mask and interpolated to 1mm isotropic; (iv) Modified k-means clustering is run on both the voxel location values and the FODs to obtain spatially contiguous labels.

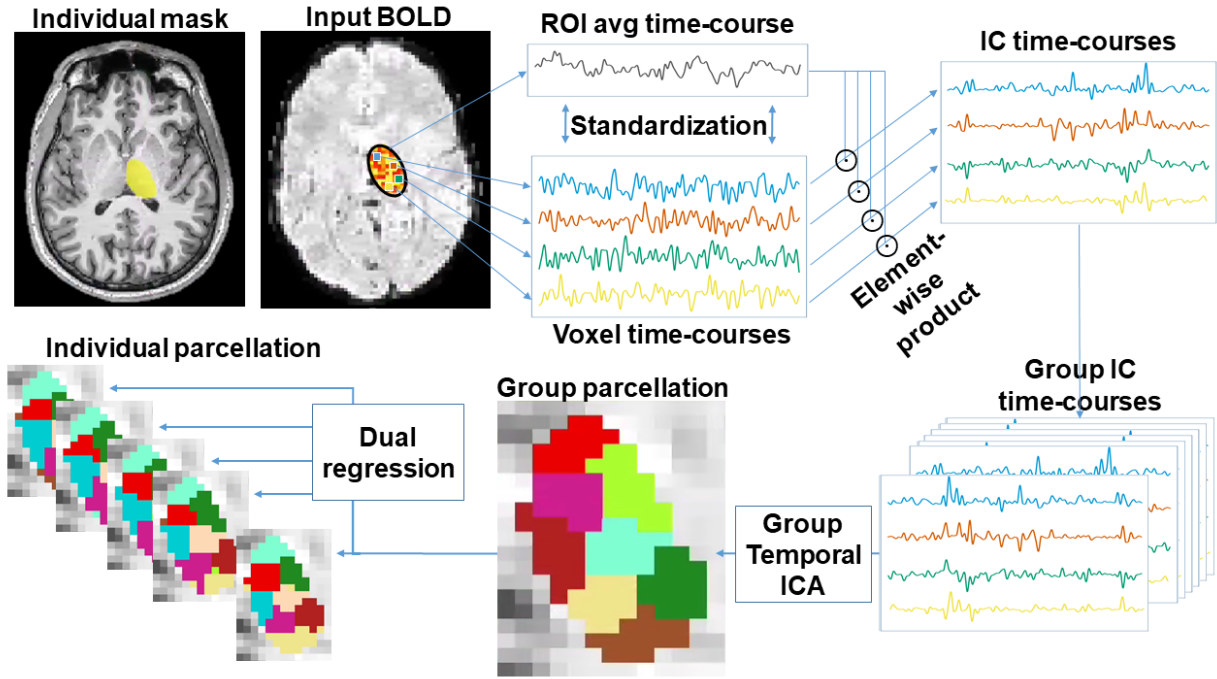

**Fig. 3** Schematic of the BOLD resting state (fMRI) processing [as described in van Oort et al., 2018]. (i) Individual thalamic mask is created via single subject segmentation of the T1-weighted MP-RAGE data; (ii) the average time-course across all voxels of thalamus (for each subject and hemisphere separately) is extracted and standardized; (iii) each individual voxel's time-course is also extracted, standardized, and then multiplied, element-wise, with the average standardized time-course; (iv) these time-courses of instantaneous correlations from all subjects are entered into a group temporal ICA with a set solution dimensionality of 30 components; (v) the resulting group parcellation is then employed to obtain, *via* dual regression, single-subject parcellation.

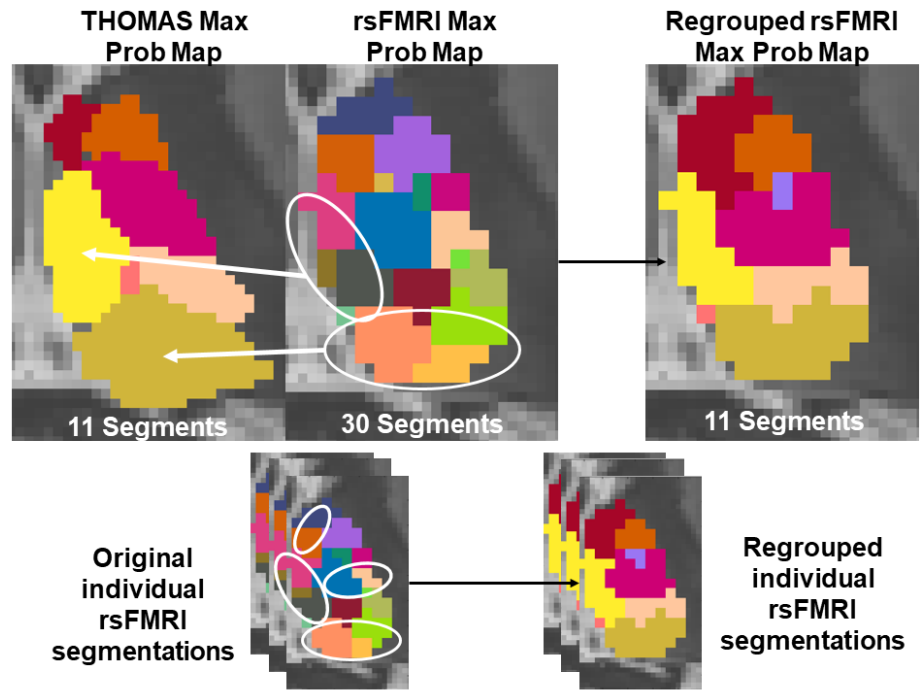

**Fig. 4** Regrouping of fMRI segmentation clusters. (i) Individual fMRI segmentation are produced with 30 clusters; (ii) fMRI and THOMAS segmentation are warped to a MNI space and maximum label probability maps are created for each method; (iii) Each of the 30 fMRI maximum probability map clusters are mapped to a THOMAS cluster on the basis of overlap; (iv) This mapping is applied at the individual level to obtain regrouped fMRI segmentation with 11 clusters.

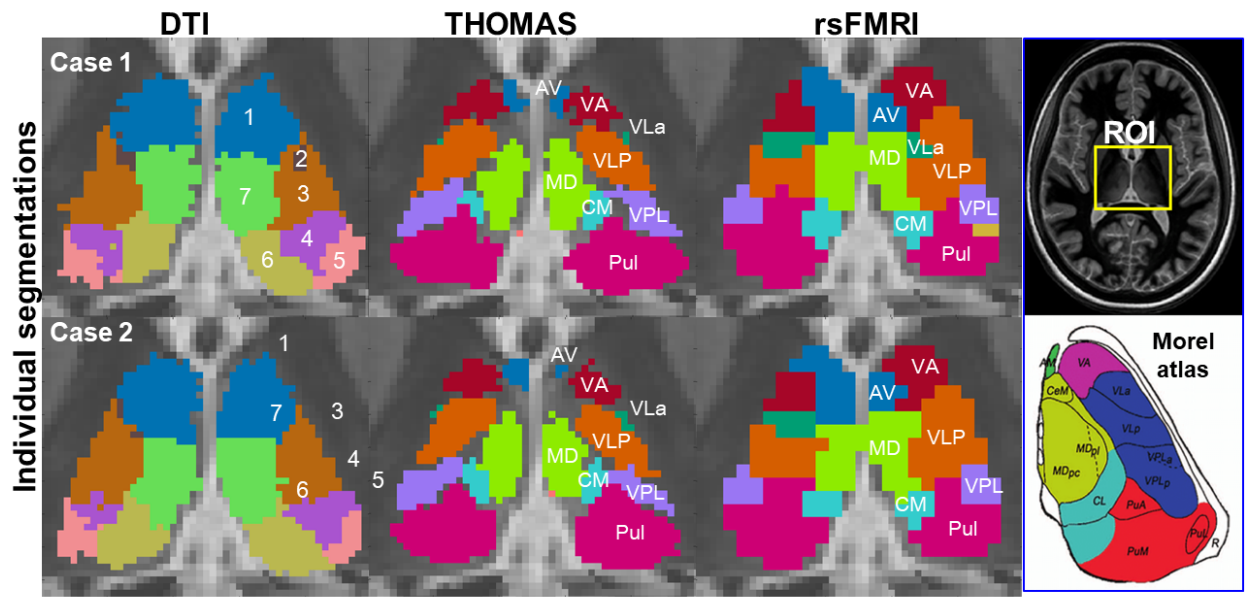

**Fig. 5** Axial view of individual thalamic segmentation for two individuals (cases 1 and 2) via structural, DTI, and fMRI based parcellation.

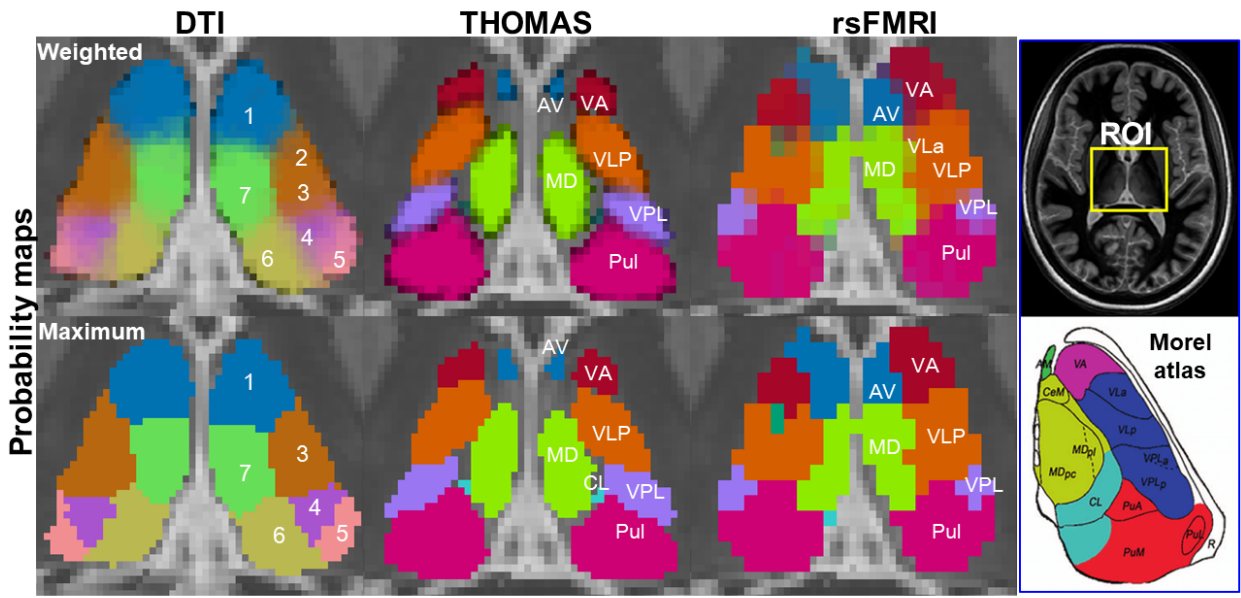

**Fig. 6** Axial, coronal, and sagittal views of weighted label probability maps for structural, DTI, and fMRI based segmentation.

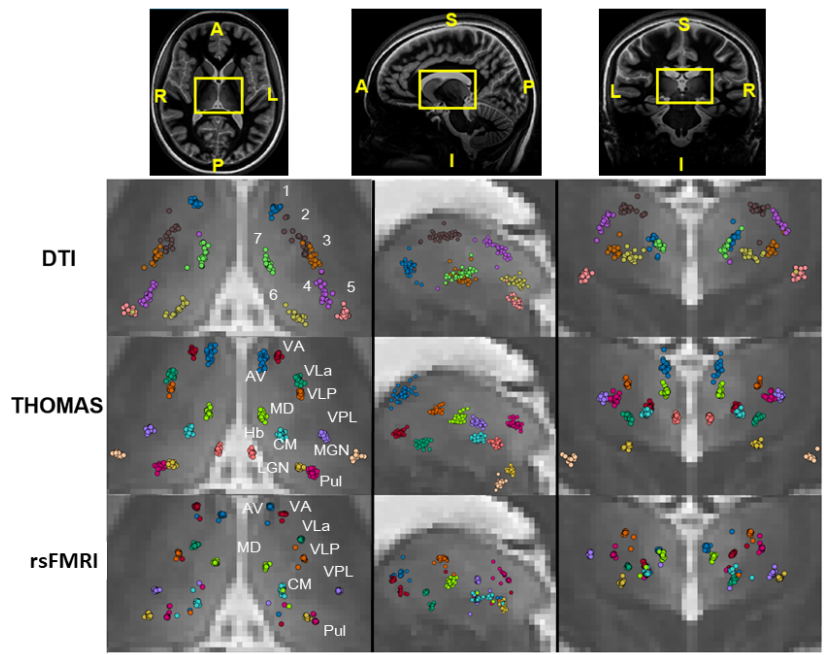

**Fig. 7** Label centroid plots in axial, sagittal, and coronal planes for structural, DTI, and fMRI parcellation methods

**Supplemental material**

| Nucleus | Dice, DTI vs. fMRI (L) | Dice, DTI vs. fMRI (R) | VSI, DTI vs. fMRI (L) | VSI, DTI vs. fMRI (R) |
|---|---|---|---|---|
| VL  | 0.42 ±0.08 | 0.44 ±0.06 | 0.75 ±0.09 | 0.85 ±0.20 |
| Pul | 0.44 ±0.07 | 0.38 ±0.07 | 0.85 ±0.08 | 0.91 ±0.21 |
| Md  | 0.51 ±0.14 | 0.47 ±0.09 | 0.93 ±0.23 | 0.50 ±0.15 |
| VPL | 0.37 ±0.11 | 0.22 ±0.07 | 0.68 ±0.17 | 0.76 ±0.11 |
| VA  | 0.37 ±0.05 | 0.43 ±0.12 | 0.93 ±0.14 | 0.42 ±0.07 |
| AV  | 0.20 ±0.06 | 0.35 ±0.09 | 0.49 ±0.13 | 0.84 ±0.08 |
| CM  | 0.21 ±0.06 | 0.31 ±0.05 | 0.52 ±0.16 | 0.00 ±0.00 |
| LGN | 0.00 ±0.00 | 0.00 ±0.00 | 0.00 ±0.00 | 0.49 ±0.14 |
| VLa | 0.18 ±0.07 | 0.16 ±0.05 | 0.46 ±0.14 | 0.83 ±0.08 |
| MGN | 0.20 ±0.07 | 0.27 ±0.08 | 0.40 ±0.12 | 0.95 ±0.11 |
| Hb  | 0.00 ±0.00 | 0.00 ±0.00 | 0.00 ±0.00 | 0.00 ±0.00 |

**Supplemental Table 1** Mean +/- standard deviation Dice scores between left and right DTI and fMRI thalamus parcellation across 18 subjects for 11 nuclei. Nuclei are arranged in descending order of average label volume.

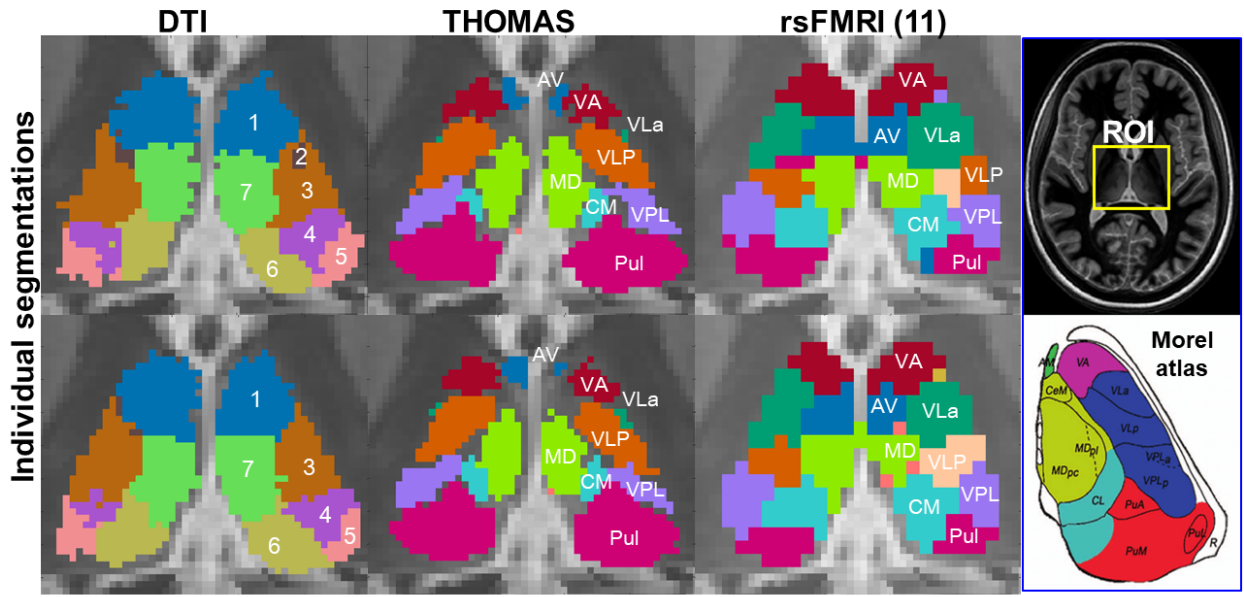

**Fig. 1** Comparison of individual thalamic segmentation for two subjects (case 1 and case 2) in axial plane. Pictured are three techniques: structural segmentation (THOMAS), fMRI segmentation reduced to 11 clusters from 30 via regrouping, and fMRI segmentation using 11 segments directly.

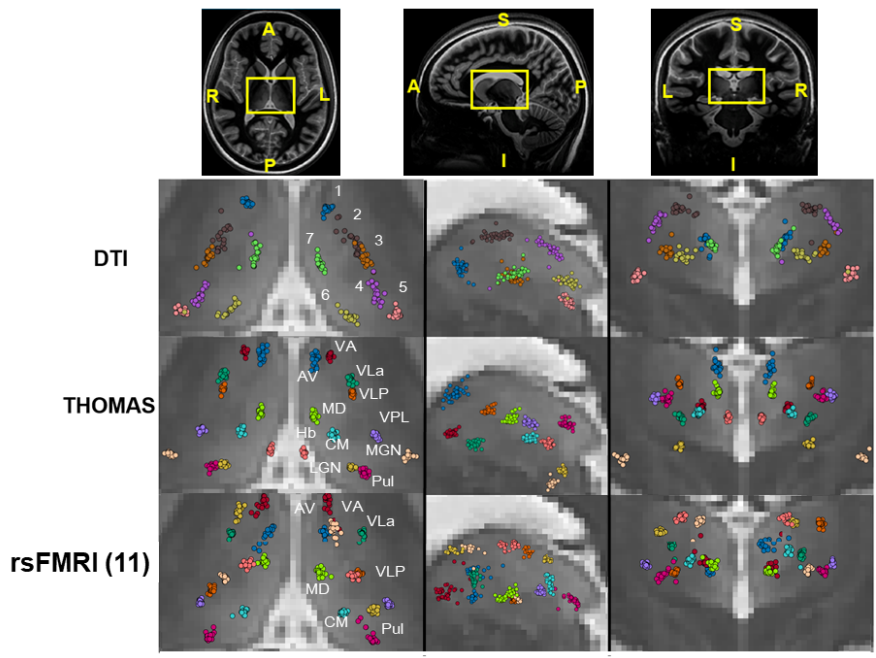

**Fig. 2** Axial view of weighted and maximum probability maps for structural segmentation (THOMAS), fMRI segmentation reduced to 11 clusters from 30 via regrouping, and fMRI segmentation using 11 segments directly